\documentclass[conference]{IEEEtran}
\IEEEoverridecommandlockouts
\usepackage{cite}
\usepackage{amsmath,amssymb,amsfonts}
\usepackage{algorithmic}
\usepackage{graphicx}
\usepackage{textcomp}
\usepackage{xcolor}
\definecolor{mygreen}{rgb}{0.1,.6,0.1}
\usepackage{subfig}
\def\BibTeX{{\rm B\kern-.05em{\sc i\kern-.025em b}\kern-.08em
    T\kern-.1667em\lower.7ex\hbox{E}\kern-.125emX}}
    
\usepackage{tabu}

\begin{document}

\title{FPGA Implementations of Layered MinSum LDPC Decoders Using RCQ Message Passing}

\author{\IEEEauthorblockN{Caleb Terrill, Linfang Wang, Sean Chen, Chester Hulse, Calvin Kuo, Richard Wesel}
\IEEEauthorblockA{\textit{Department of Electrical and Computer Engineering} \\
\textit{University of California, Los Angeles}\\
Los Angeles, United States \\
\{cterrill26,lfwang,mistystory,chulse,calvinkuo,wesel\}@ucla.edu}
\and
\IEEEauthorblockN{Dariush Divsalar}
\IEEEauthorblockA{\textit{Jet Propulsion Laboratory} \\
\textit{Cal. Institute of Technology}\\
Pasadena, United States \\
dariush.divsalar@jpl.nasa.gov}
}

\maketitle

\begin{abstract}
Non-uniform message quantization techniques such as reconstruction-computation-quantization (RCQ) improve error-correction performance and decrease hardware complexity of low-density parity-check (LDPC) decoders that use a flooding schedule. Layered MinSum RCQ (L-msRCQ) enables message quantization to be utilized for layered decoders and irregular LDPC codes. We investigate field-programmable gate array (FPGA) implementations of L-msRCQ decoders. Three design methods for message quantization are presented, which we name the Lookup, Broadcast, and Dribble methods. The decoding performance and hardware complexity of these schemes are compared to a layered offset MinSum (OMS) decoder.  Simulation results on a (16384, 8192) protograph-based raptor-like (PBRL) LDPC code show that a 4-bit L-msRCQ decoder using the Broadcast method can achieve a 0.03 dB improvement in error-correction performance while using 12\% fewer registers than the OMS decoder. A Broadcast-based 3-bit L-msRCQ decoder uses 15\% fewer lookup tables, 18\% fewer registers, and 13\% fewer routed nets than the OMS decoder, but results in a 0.09 dB loss in performance.
\end{abstract}


\section{Introduction}
Low-density parity-check (LDPC) codes are powerful error-correcting codes that approach channel capacity when high-precision message passing is used while decoding. Realistic hardware implementations of LDPC decoders limit the width of the messages passed between variable nodes (VNs) and check nodes (CNs). This sacrifices error-rate performance to achieve lower resource utilization and wiring complexity. Uniformly quantized messages typically range from 5-7 bits, with anything less resulting in a sizable reduction in error-correction ability \cite{Jinghu_Chen2005-vu,Zhang2007-qu}. 

A significant amount of research has been put into reducing message widths to less than 5 bits. Non-uniform quantization schemes, which effectively extend the dynamic range of the quantizers, use only 3 or 4 bits to achieve similar or even better performance than full-precision belief propagation (BP) and MinSum decoding \cite{Stark2019-bm,Meidlinger2017-aa,Stark2018-ff,Lewandowsky2018-pg,Romero2016-jn,-_Lee2005-ni,He2019-hr,Wang2020-kh}. J. K. Lee \emph{et al.} propose the mutual information maximization quantized belief propagation (MIM-QBP)\cite{-_Lee2005-ni} decoder, which designs iteration-specific non-uniform quantizers and reconstruction mappings at nodes. Both VN and CN operations are simple mappings and fixed point additions for MIM-QBP.  X. He in \cite{He2019-hr} shows how to systematically design the parameters for quantization and reconstruction modules.  L. Wang {\em et al.} further generalize the MIM-QBP structure and propose a reconstruction-computation-quantization (RCQ) paradigm \cite{Wang2020-kh} which allows CNs to use the Min operation for computation reduction.

Existing works on non-uniform quantization decoders have typically focused on regular LDPC codes, which are convenient for hardware decoders due to the uniformity of the VN and CN degrees\cite{Lewandowsky2018-pg,Romero2016-jn,-_Lee2005-ni,He2019-hr}. Some attention has been given to irregular codes, which have better error-rate performance, but present challenges with message quantization and hardware design due to the varying node degrees \cite{Meidlinger2017-aa,Stark2018-ff,Stark2019-bm,Wang2020-kh,Luby2001-ut}. Additionally, a fully parallel or flooding schedule for CN operations and VN updates is generally assumed \cite{-_Lee2005-ni,He2019-hr,Lewandowsky2018-pg,Romero2016-jn,Meidlinger2017-aa}. This approach may be undesirable or even unfeasible for decoder implementations using codes with large block lengths, where the design of a fully parallel decoder becomes very resource intensive. Instead, a partially parallel or layered architecture is often preferred \cite{-_Chang2008-vk}.

\subsection{Contributions}
%-----Linfang-----
As a primary contribution, we investigate field-programmable gate array (FPGA) implementations of a layered MinSum RCQ (L-msRCQ) decoder for irregular LDPC codes. Three design methods are presented in this paper:
\begin{itemize}
    \item \emph{Lookup method}: all L-msRCQ parameters are stored in local read-only memories (ROMs) in VNs. Reconstruction and quantization are performed as ROM lookups.
    \item \emph{Broadcast method}: L-msRCQ parameters are stored in a centralized ROM and broadcasted to VNs. Quantization and reconstruction are performed using small lookup tables (LUTs) as logic.
    \item \emph{Dribble method}: the current layer's L-msRCQ parameters are stored locally at VNs in registers, quantization and reconstruction are performed using LUTs as logic.
\end{itemize}
We use these methods to design 3-bit and 4-bit L-msRCQ decoders for a quasi-cyclic (QC) protograph-based raptor-like (PBRL) LDPC code\cite{Chen2015-ul} with code length 16384 and rate $0.5$. The decoding performance and hardware complexity of these three methods are analyzed. As a reference, a series of layered offset MinSum (OMS) decoders using uniform message quantization of 5-7 bits are provided. Simulation and hardware utilization results show that the 4-bit L-msRCQ decoder using the Broadcast method can achieve a 0.03 dB increase in error-correction performance and a 12\% decrease in register usage compared to the OMS reference decoder. A 3-bit Broadcast-based L-msRCQ decoder results in a 0.09 dB loss in performance, but requires 15\% fewer LUTs, 18\% fewer registers, and 13\% fewer routed nets than the OMS decoder.

\subsection{Outline}

The remainder of this paper is organized as follows. Section II provides background on layered MinSum decoding, OMS decoding, and RCQ. Section III presents the reference OMS decoder architecture. Section IV describes the Lookup, Broadcast, and Dribble methods used for the L-msRCQ implementations. Section V presents the frame error rate (FER) performance and resource utilization of each decoding scheme.
\section{Background}

An LDPC code is defined by a sparse $M \times N$ parity check matrix $H$, where $M$ is the number of CNs and $N$ is the number of VNs.
%If $H_{mn} = 1$, then variable $n$ is part of CN operation $m$.
%Let $N(m)$ denote the set of variables that contribute to CN operation $m$, and let $M(n)$ represent the set of CN operations that variable $n$ is a part of. 
Let $N(m)$ denote the set of VNs that connect to CN $m$, let $M(n)$ represent the set of CNs that connect to VN $n$, and let $i$ stand for the decoding iteration. We define the following notations:

\begin{itemize}
\item $U_{ch,n}$: the log-likelihood ratio (LLR) for variable $n$ derived from the channel output.
\item $U_{mn}^i$: the LLR message passed from CN $m$ to VN $n$.
\item $V_{mn}^i$: the LLR message passed from VN $n$ to CN $m$.
\item $V_n$ : the a posteriori LLR (AP-LLR) of variable $n$.
\end{itemize}

\subsection{Layered MinSum and Offset MinSum Decoding}
The MinSum and OMS algorithms for LDPC decoding are approximations of the BP algorithm, offering simplified CN operations at a small performance loss \cite{Fossorier1999-bp,Chen2002-xi}. A horizontal layered decoding approach processes sets of CNs sequentially, enabling better code performance compared to a flooding approach with respect to iterations required, while using less hardware resources \cite{Hocevar2004-xi}. We consider decoding iterations $i = 1,2,3, \dots ,I_{max}$. To begin a decoding iteration, the AP-LLRs are initialized:
\begin{equation}
V_n=U_{ch,n}\label{Layered VN Init1}
\end{equation}
A VN-to-CN message is calculated:
\begin{equation}
V_{mn}^i=V_n-U_{mn}^{i-1} \label{Layered VN-to-CN}
\end{equation}
For the MinSum algorithm, CN-to-VN messages are obtained:
\begin{equation}
U_{mn}^i=\left(\prod_{n' \in N(m)\backslash n}\text{sgn}(V_{mn'}^i)\right)\times \min_{n' \in N(m)\backslash n}|V_{mn'}^i|\label{MS CN-to-VN}
\end{equation}
The OMS algorithm subtracts a small positive constant $\alpha$ from the CN-to-VN messages for a better approximation of BP:
\begin{equation}
\begin{aligned}
U_{mn}^i= & \left(\prod_{n' \in N(m)\backslash n}\text{sgn}(V_{mn'}^i)\right)\times \\
          & \max\left(\min_{n' \in N(m)\backslash n}|V_{mn'}^i| - \alpha, 0\right)\label{OMS CN-to-VN}
\end{aligned}
\end{equation}
The AP-LLRs are updated:
\begin{equation}
V_n=V_{mn}^i+U_{mn}^i \label{Layered VN}
\end{equation}
%Hard decisions after iteration $i$ are reached with the rule:
%\begin{equation}
%Z_n^i = \begin{cases} 
%      0 & V_n \geq 0\\
%      1 & V_n < 0
%   \end{cases}\label{MS Decision}
%\end{equation}
Decoding terminates after the syndrome check is passed or $I_{max}$ iterations have taken place. 

For hardware decoder implementations using message quantization, the bit widths of the VN-to-CN messages calculated in Eq.\eqref{Layered VN-to-CN} are reduced. We refer to the quantized values as $\hat{V}_n^{im}$. The CN operations in Eq.\eqref{MS CN-to-VN} or Eq.\eqref{OMS CN-to-VN} operate on these reduced precision values to calculate the quantized CN-to-VN messages $\hat{U}_n^{im}$. In this paper, we use the pair $(b^c,b^v)$ to describe a decoder's bit width, where $b^c$ denotes the bit width for CN operations and $b^v$ represents the bit width for $V_n$.

%Equations \eqref{Layered VN-to-CN} and \eqref{Layered VN} describe how CNs can be processed one at a time. This implies that a layer consists of a single CN. In practical hardware decoder implementations targeting a balance between throughput and hardware efficiency, the sequential processing of CNs limits throughput too drastically. To achieve a higher speed layered decoder, layers must consist of multiple CNs. 

QC-LDPC codes are a family of structured LDPC codes characterized by a parity check matrix which consists of $L\times L$ square sub-matrices that are either the zero matrix or cyclic permutations of the identity matrix (also called a circulant). By grouping $L$ CNs that form a row of circulants into a layer, each variable connects to at most one CN per layer. This allows the VN-to-CN message calculations and VN update rules from Eq.\eqref{Layered VN-to-CN} and Eq.\eqref{Layered VN} to be applied to all the CNs and associated VNs in a layer at once, without needing to account for the same VN being updated multiple times. As a result, a total of $L$ CNs can be processed at once, achieving an excellent balance between decoder throughput and hardware complexity.

%Layered LDPC decoders typically use codes with a QC parity matrix. A QC-LDPC parity matrix consists of $L \times L$ blocks, where each block is either a rotated identity matrix or a zero matrix. %The notation $I_a$ denotes an $L \times L$ identity matrix, where each row has been cyclically shifted to the right by $a$ positions. We define $I_{\infty}$ to be the zero matrix. A QC parity matrix can thus be described by its values of $a$ for each block.
%\begin{equation}
%H_a = \begin{bmatrix}
%    a_{11} & a_{12} & \dots  & a_{1\frac{N}{L}} \\
%    a_{21} & a_{22} & \dots  & a_{2\frac{N}{L}} \\
%    \vdots & \vdots & \ddots & \vdots \\
%    a_{\frac{M}{L}1} & a_{\frac{M}{L}2} & \dots  & a_{\frac{M}{L}\frac{N}{L}}
%\end{bmatrix} \label{H Matrix}
%\end{equation}
%This leads to the natural grouping of CNs that form a row of blocks into a layer. By choosing layers in this fashion, each variable is part of at most one CN operation per layer. This allows the VN-to-CN message calculations and VN update rules from \eqref{Layered VN-to-CN} and \eqref{Layered VN} to be applied to all the CNs and associated VNs in a layer at once, without needing to account for the same VN being updated multiple times. Using this scheme, a total of $L$ CNs can be processed at once, achieving an excellent balance between decoder throughput and hardware complexity.
\subsection{Reconstruction-Computation-Quantization Decoder}
An RCQ decoder allows for a low bit width representation of $\hat{V}^{i}_{mn}$ and $\hat{U}^{i}_{mn}$ by setting dynamic quantization and reconstruction mappings. The quantizers $Q^{(i,l)}(\cdot)$ compress a $b^v$-bit message $V^{i}_{mn}$ to a $b^c$-bit message $\hat{V}^{i}_{mn}$, and reconstructions $R^{(i,l)}(\cdot)$ map a $b^c$-bit message  $\hat{U}^{i}_{mn}$ to a $b^{v}$-bit message ${U}^{i}_{mn}$. The superscripts $i$, $l$ represent the iteration number and layer index. More details on RCQ parameters are seen in \cite{Wang2020-kh}. In this work, we are focused on the L-msRCQ decoder, where layers are processed sequentially and Eq.\eqref{MS CN-to-VN} is performed at the CNs. We refer to L-msRCQ and RCQ interchangeably for the remainder of this paper. 

\section{OMS Decoder Architecture}
This section presents our FPGA design for the layered OMS decoder, which can be extended to implement the L-msRCQ decoder. The structure of our layered OMS decoder can be broken into three main components: VN banks, a CN pipeline, and a control unit. The architecture is largely governed by the construction of the chosen LDPC code. Since the parity matrix for a QC-LDPC code has an $L \times L$ QC structure, the CN pipeline must be able to process the operations for $L$ CNs in parallel. Additionally, a total of $L$ VN banks are necessary, each able to send and receive one message from the CN pipeline per cycle.

\begin{figure}[t]
	\centering
	\includegraphics[width=20pc]{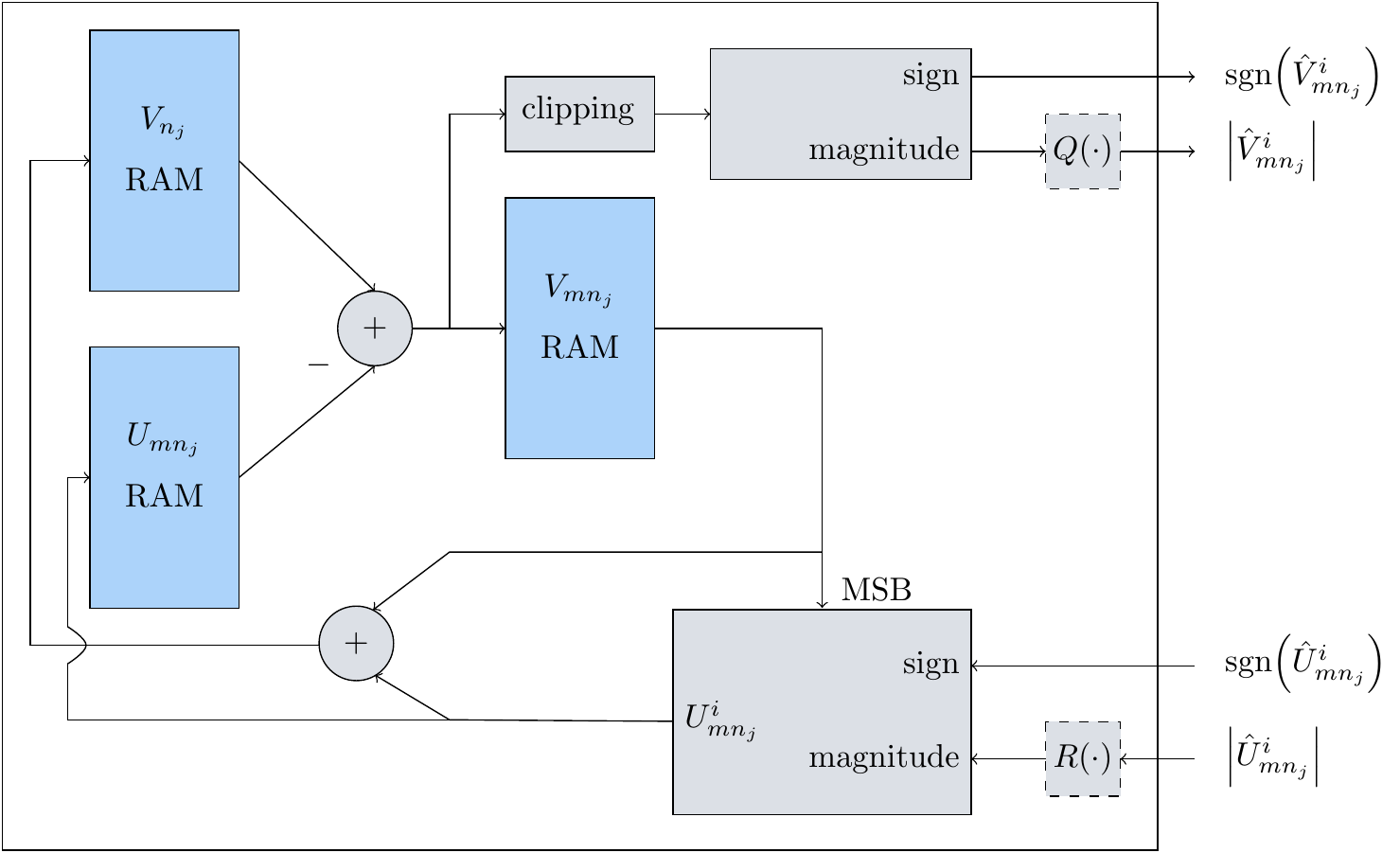}
	\caption{ VN Bank Block Diagram.}
    \label{fig}
\end{figure}
\subsection{VN Bank}

The structure of a VN bank is shown in Fig. 1. A single VN bank is responsible for $\frac{N}{L}$ variables. VN bank $j$ contains variables $n_j = \{j, L+j, 2L+j, \dots ,(\frac{N}{L}-1)L + j\}$. This ensures that variables which share the same circulants of the parity matrix are partitioned to different VN banks. Messages from all the variables in a circulant can be simultaneously sent and received by the $L$ total VN banks and the CN pipeline. 

The $V_{n_j}$ RAM, $U_{mn_j}$ RAM, and $V_{mn_j}$ RAM in Fig. 1. each allow for one read and one write per cycle. To calculate  $V^{i}_{mn}$, the required values from the $V_{n_j}$ RAM and $U_{mn_j}$ RAM can be read out and subtracted. This result is sent to the CN pipeline and temporarily stored in the $V_{mn_j}$ RAM. $V^{i}_{mn}$ is clipped to a $b^c$-bits message before being forwarded to the CN pipeline:
\begin{equation}
\hat{V}_n^{im} = \begin{cases} 
      2^{b^c-1}-1 & V_n^{im} > 2^{b^c-1}-1\\
      V_n^{im} & -2^{b^c-1} \leq V_n^{im} \leq 2^{b^c-1}-1\\
      -2^{b^c-1} & V_n^{im} < -2^{b^c-1}
   \end{cases}\label{Clipping}
\end{equation}
%This clipping rule treats all message bits as integer bits.
The sign and magnitude of $\hat{V}_n^{im}$ are sent to the CN pipeline separately. Note that $Q(\cdot)$ and $R(\cdot)$ are not included in the OMS decoder.

\subsection{CN pipeline}
To process a layer, the VN-to-CN messages for every non-zero circulant in the layer are sequentially calculated and sent to the CN pipeline. The structure of the CN pipeline is shown in Fig. 2. The CN pipeline receives $L$ messages per cycle from the $L$ VN banks. Circular shifters align each wave of incoming messages with the CNs they contribute to. For a circulant in the parity matrix that is the identity matrix cyclically permuted by $p$, its $L$ total VN-to-CN messages will be circularly shifted by $p$ so that the messages contribute to the correct CNs.
%, A set of messages from a circulant are cyclically shifted by the amount the same amount as the circulant 
%\textit{Barrel shifters align the incoming messages with the CNs they contribute to according to the parity matrix, where each shift is by the amount $a$ associated with the block}{\color{red} This sentences is no clear, can you rewrite it ? Thanks!}. 
The aligned values are fed into logic which calculates the first minimum (MIN1) magnitude and second minimum (MIN2) magnitude of the input messages, along with an XOR of all the sign bits (SIGN). 
%Once all messages from a layer's non-zero blocks have made their way to the MIN1, MIN2, and SIGN calculation, these values are ready to be returned to the VN banks. 

CN-to-VN messages are sequentially returned to the VN banks for each of the layer's non-zero circulants. Along with the SIGN, the MIN1 or MIN2 value of a CN is sent back to its corresponding variables, depending on whether a VN provided the MIN1 value. An offset is subtracted from the selected MIN1 or MIN2 value. Circular shifters realign the MIN1 or MIN2 and SIGN values with the correct VN banks.

Each VN bank receives a sequence of CN-to-VN messages as sign and magnitudes values. When a message arrives, the corresponding $V^{i}_{mn}$ message for the variable is read from the $V_{mn_j}$ RAM. The sign bit of $V^{i}_{mn}$ is XORed with the incoming sign value, and the result is used to calculate the signed CN-to-VN message $U^{i}_{mn}$. The result is stored in the $U_{mn_j}$ RAM, and it is summed with $V^{i}_{mn}$ to form an updated AP-LLR which is stored in the $V_{n_j}$ RAM. To allow the AP-LLR values to grow, the $V_{n_j}$ RAM stores $b^v = b^c + 2$ bits.
\begin{figure}[t]
	\centering
	\includegraphics[width=20pc]{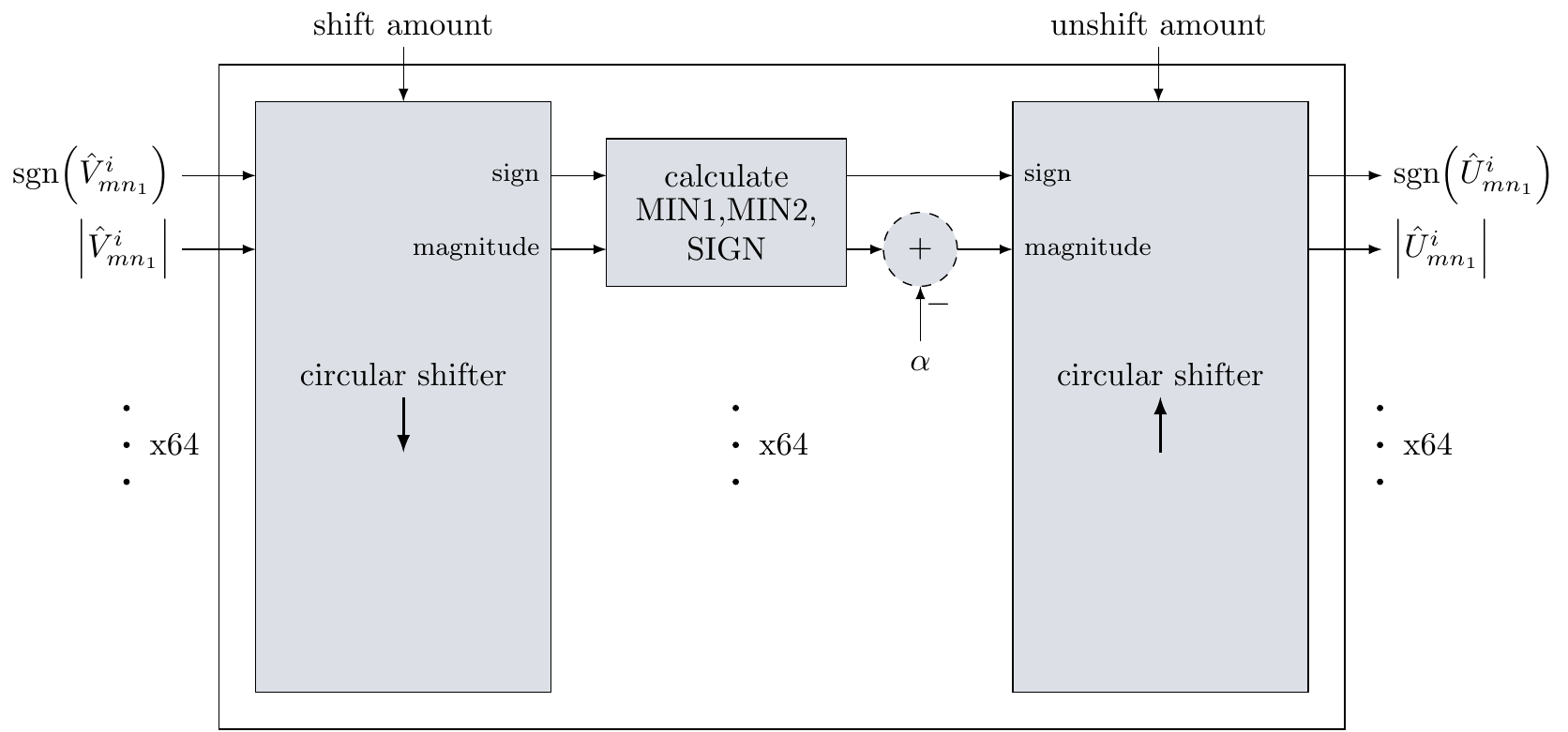}
	\caption{ CN pipeline Diagram.}
    \label{fig}
\end{figure}
\subsection{Control Unit}
The control signals that dictate the logic of the VN banks and CN pipeline are provided by the control unit. A state machine manages the readout of values from VN bank RAMs, the shift amounts for the CN pipeline, and the writing of results back into the VN bank RAMs. To allow for a high decoder throughput, the described datapath is pipelined into multiple stages to enable a high frequency clock. The control unit overlaps the processing of adjacent layers, meaning that after the data for layer $l$ has been read out from the VN banks and sent to the CN pipeline, the values for layer $l+1$ are immediately read out before the writeback of layer $l$ completes. This allows for high utilization of the datapath pipeline, but creates read after write (RAW) hazards for VNs that are contained in consecutive layers, since the updated AP-LLR for a variable may not be written by the time it needs to be read out by the next layer. To overcome this, the sequential order in which the control unit reads circulants out of the VN banks has been hand modified so that VNs which encounter RAW hazards are read earlier or later in their layer's sequence to prevent overlap from occurring.

\section{L-msRCQ Decoder Architecture}

The L-msRCQ decoder FPGA design expands on the design of the layered OMS decoder. The key differences between the L-msRCQ and layered OMS decoders are:
\begin{itemize}
    \item The L-msRCQ decoder requires the $Q(\cdot)$ and $R(\cdot)$ functions in Fig. 1 to quantize and reconstruct message magnitudes.
    \item Because the MinSum algorithm is used for CN operations, the offset calculation in Fig. 2 is not included in the CN pipeline.
\end{itemize}
 To implement the quantization and reconstruction operations in the VN banks, we have devised the \emph{Lookup}, \emph{Broadcast}, and \emph{Dribble} methods. These three approaches are functionally identical, but the way in which they compute the $Q(\cdot)$ and $R(\cdot)$ operations differ, resulting in unique hardware utilizations. 

\subsection{Lookup Method}
Since the quantization and reconstruction functions simply map an input message to an output message, an implementation based on a lookup into a ROM is evident. To quantize $ |V^{i}_{mn}|$, the three-tuple $( |V^{i}_{mn}|, i, l)$ is used to index into a ROM to obtain $|\hat{V}^{i}_{mn}|$. Similarly for $\hat{U}_{mn}^i$, the three-tuple $( |\hat{U}^{i}_{mn}|, i, l)$ forms an address to index into a ROM to read out $|U^{i}_{mn}|$. These $Q(\cdot)$ and $R(\cdot)$ functions in every VN bank require their own ROMs, implemented using block RAMS (BRAMs). Assuming one BRAM is used for $Q(\cdot)$ and one is used for $R(\cdot)$, then $L$ VN banks with two ROMs each results in a total of $2L$ additional BRAMs used. If BRAMs with multiple ports are available, then they can be shared by different VN banks to reduce the total amount required. 

% If no ROM sharing occurs, then $L$ VN banks with two ROMs each results in a total of $2L$ additional BRAMs used. This amount can be reduced with multiport ROMs and other synthesis techniques.

\subsection{Broadcast Method}

The Broadcast method provides a scheme where L-msRCQ parameters are centralized, instead of being stored in each VN bank. The pair $(i,l)$ is used to index into ROMs in the control unit. These ROMs output quantization thresholds $th^{(i,l)}_1, th^{(i,l)}_2, \dots , th^{(i,l)}_{2^{b^c-1}-1}$ and reconstruction values $re^{(i,l)}_1, re^{(i,l)}_2, \dots , re^{(i,l)}_{2^{b^c-1}}$, which are wired to the VN banks. The $Q(\cdot)$ and $R(\cdot)$ blocks in the VN banks take in the parameters and use logic to perform their respective operations. Fig. 3 shows an implementation for a 3-bit L-msRCQ, which uses a mere 2 bits for quantized message magnitudes. Two BRAMS are required in the control unit for the quantization thresholds and reconstruction values. Logic implemented using small LUTs is needed in the VN units because of the $Q(\cdot)$ and $R(\cdot)$ functions. The main penalty comes with the wiring necessary to route the L-msRCQ parameters from the control unit to VN banks. As shown in Fig. 3, if $w$ bits are used for each of the thresholds and reconstruction values of 3-bit L-msRCQ, a total of $7w$ additional wires need to be routed to a VN bank. With $L$ VN banks, the total amount of added routes increases by $7wL$. For a 4-bit L-msRCQ decoder, the total increase is $15wL$. It is to be noted that the same parameters are routed to all the VN banks, meaning that the required wiring can be reduce by path sharing.

\begin{figure}[t] 
    \centering
  \subfloat[Quantization logic\label{8k_fer}]{%
       \includegraphics[width=11pc]{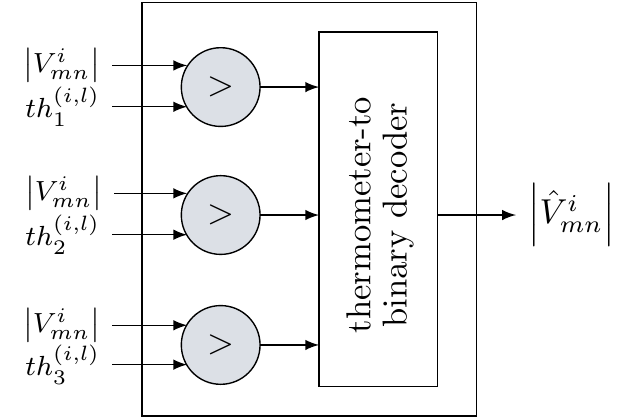}}
    \\
  \subfloat[Reconstruction logic\label{8k_adit}]{%
        \includegraphics[width=11pc]{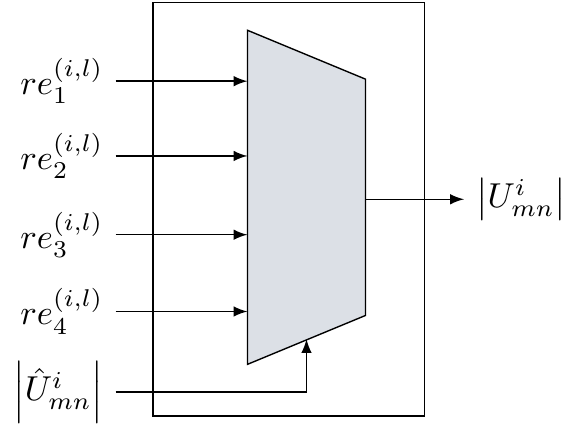}}
  \caption{Quantization and Reconstruction Logic}
\end{figure}

\subsection{Dribble Method}
The Dribble method attempts to reduce the number of long wires required by the Broadcast method to send L-msRCQ parameters from the control unit to the VN banks. Registers in the VN banks save the current thresholds and reconstruction values necessary for the $Q(\cdot)$ and $R(\cdot)$ functions. Once again, quantization and reconstruction can be implemented using the logic in Fig. 3. When a new set of parameters is required, the bits are transferred one by one or in small batches from the control unit to the VN bank registers. Just as in the Broadcast method, two extra BRAMs and logic for the $Q(\cdot)$ and $R(\cdot)$ functions are required. But where the Broadcast method needs $7w$ additional wires routed to each VN bank for 3-bit L-msRCQ, the Dribble method requires only as many wires as the transfer batch size. The penalty of the Dribble method comes with the extra usage of registers in the VN banks. A total of $7w$ bits stored in registers would be necessary in each VN bank to save the current threshold and reconstruction values for 3-bit L-msRCQ. In total, $7wL$ bits of register storage would be used for 3-bit L-msRCQ, and $15wL$ bits would be necessary for 4-bit L-msRCQ. This total can be reduced by having multiple VN banks share sets of registers.

\section{FER and Resource Utilization Results}

\begin{figure}[t]
\centerline{\includegraphics[scale=0.6]{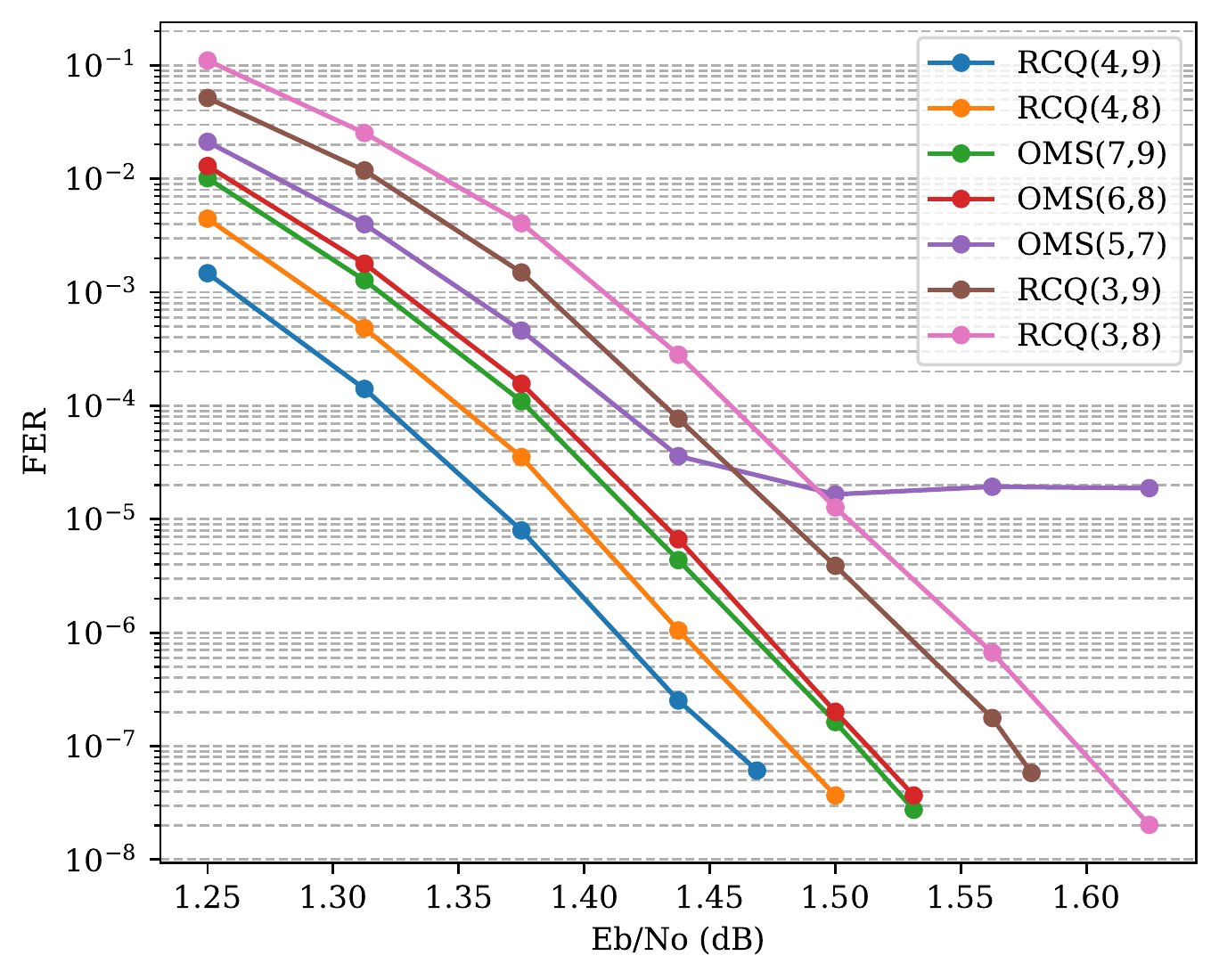}}
\caption{FER vs. $\frac{E_b}{N_o}$ (dB) for OMS and L-msRCQ Decoders.}
\label{fig}
\end{figure}

To test the OMS decoder and the Lookup, Broadcast, and Dribble L-msRCQ methodologies, the designs have been implemented on the programmable logic of a Xilinx Zynq UltraScale+ MPSoC device. Each design met timing with a 500 MHz clock. The maximum decoding iterations, $I_{max}$, is set to 16, and for the OMS decoders we set $\alpha = 0.5$.   %Since the datapath pipelines for all the decoders have a nearly identical cycle latency, the throughputs for the designs are almost identical. When the lowest rate PBRL code of $0.5$ is used, the minimum throughput when $I_{max} = 16$ decoding iterations are used is approximately 275 Mbps.
%\textcolor{red}{When the highest rate PBRL code of $0.89$ is used with a frame error rate  less than $10^{-5}$, the throughput is approximately 1.7 Gbps. } 
% (160 Mbps)  (320 Mbps)

\subsection{FER Performance}

The FER performance for the decoder configurations at varying $\frac{E_b}{N_o}$ is shown in Fig. 4. The bit width of each decoder is specified by its $(b^c,b^v)$ pair. The RCQ(4,9) and RCQ(4,8) designs show a 0.05 and 0.03 dB improvement in FER performance respectively compared to the OMS(6,8) decoder at FER $10^{-7}$. Meanwhile, the RCQ(3,9) and RCQ(3,8) designs show a 0.06 and 0.09 dB decrease in FER performance respectively compared to OMS(6,8) at FER $10^{-7}$, while avoiding the error floor above FER $10^{-5}$ faced by the OMS(5,7) decoder. The OMS(7,9) and OMS(6,8) decoder performances are nearly identical.

\subsection{LUTs and Registers}

The LUT and register usage of the investigated decoders is shown in Fig. \ref{fig: register_LUT}. The 3-bit L-msRCQ decoders show a decrease in LUT and register usage compared to the OMS(6,8) decoder due to fewer bits being used in the CN pipeline. For the 4-bit L-msRCQ decoders, register usage decreases, but LUT count increases for the non-Lookup decoding schemes due to the logic necessary for the $Q(\cdot)$ and $R(\cdot)$ functions. Dribble-based decoders do not exhibit the expected penalty of increased register utilization, since the synthesis tool recognizes and removes the redundant registers held in different VN banks. To meet more stringent timing constraints, more of these redundant registers can be left present in the design to allow for better localization of L-msRCQ parameters to $Q(\cdot)$ and $R(\cdot)$ logic.

\subsection{BRAMs and Dynamic Power}
\begin{figure}[t] 
    \centering
  \subfloat[LUT and Register Usage\label{fig: register_LUT}]{%
       \includegraphics[width=20pc]{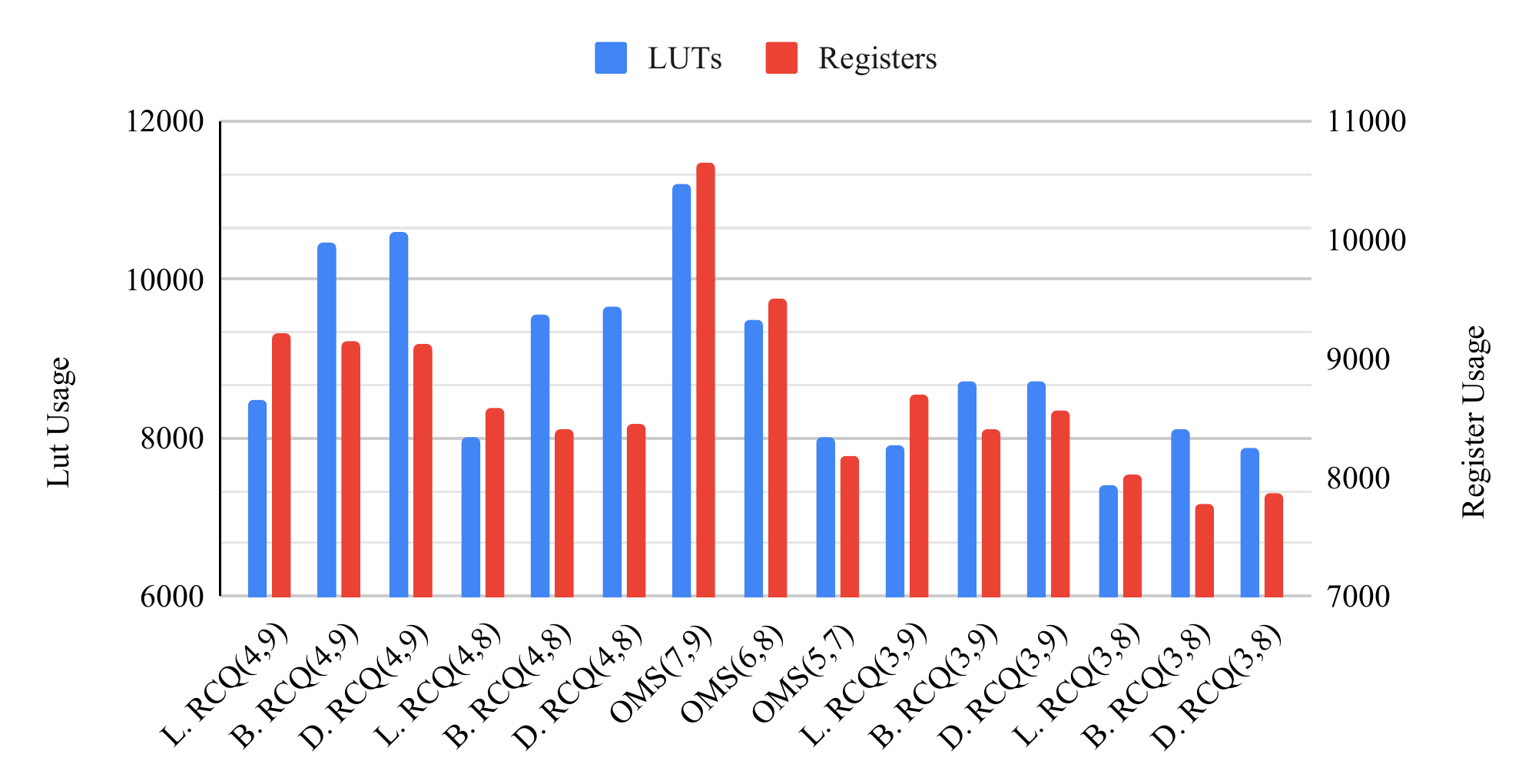}}
    \\
  \subfloat[Block RAM and Dynamic Power Usage\label{fig: bram_power}]{%
        \includegraphics[width=20pc]{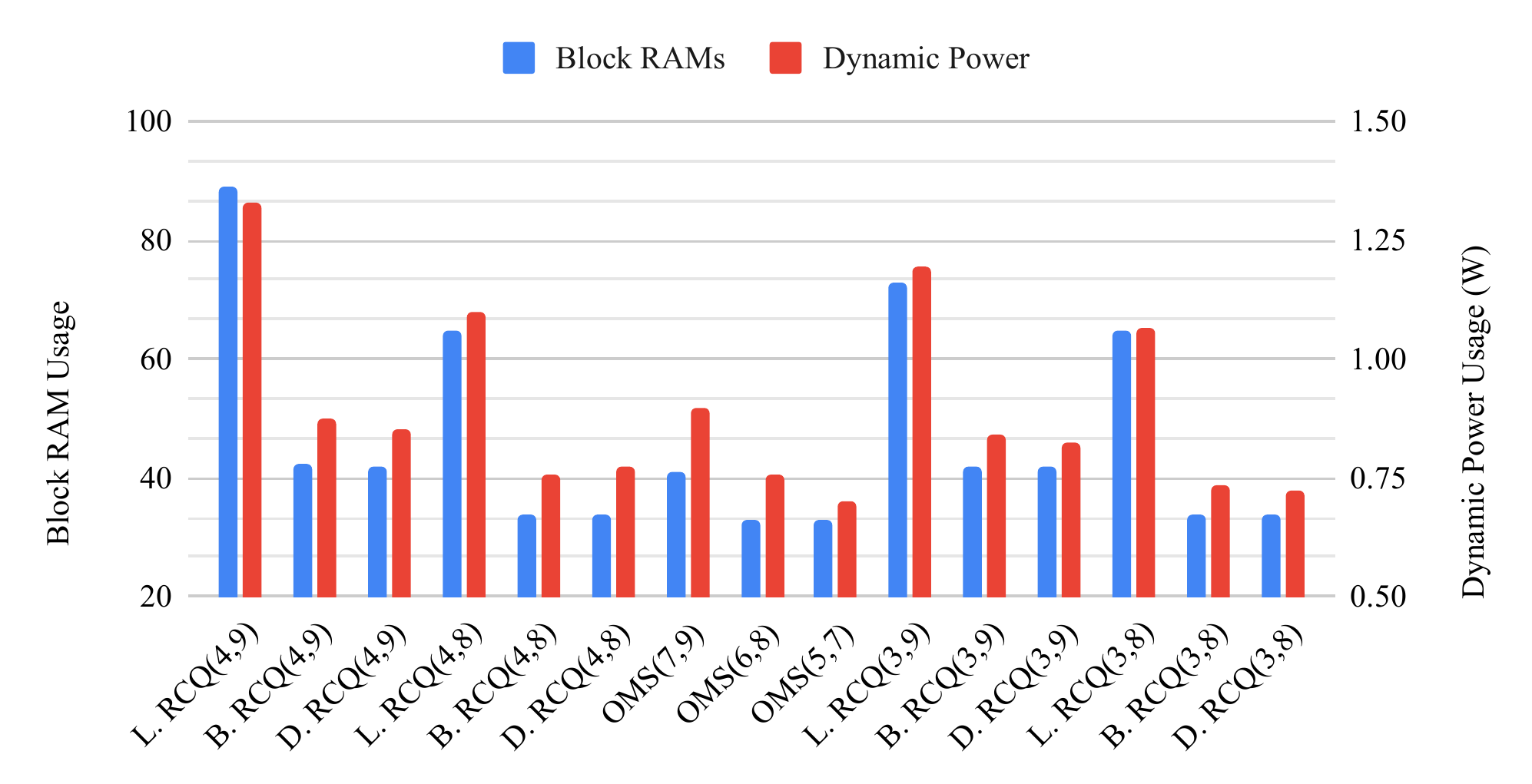}}
    \\
     \subfloat[Routed Net Usage\label{fig: rnet}]{%
        \includegraphics[width=20pc]{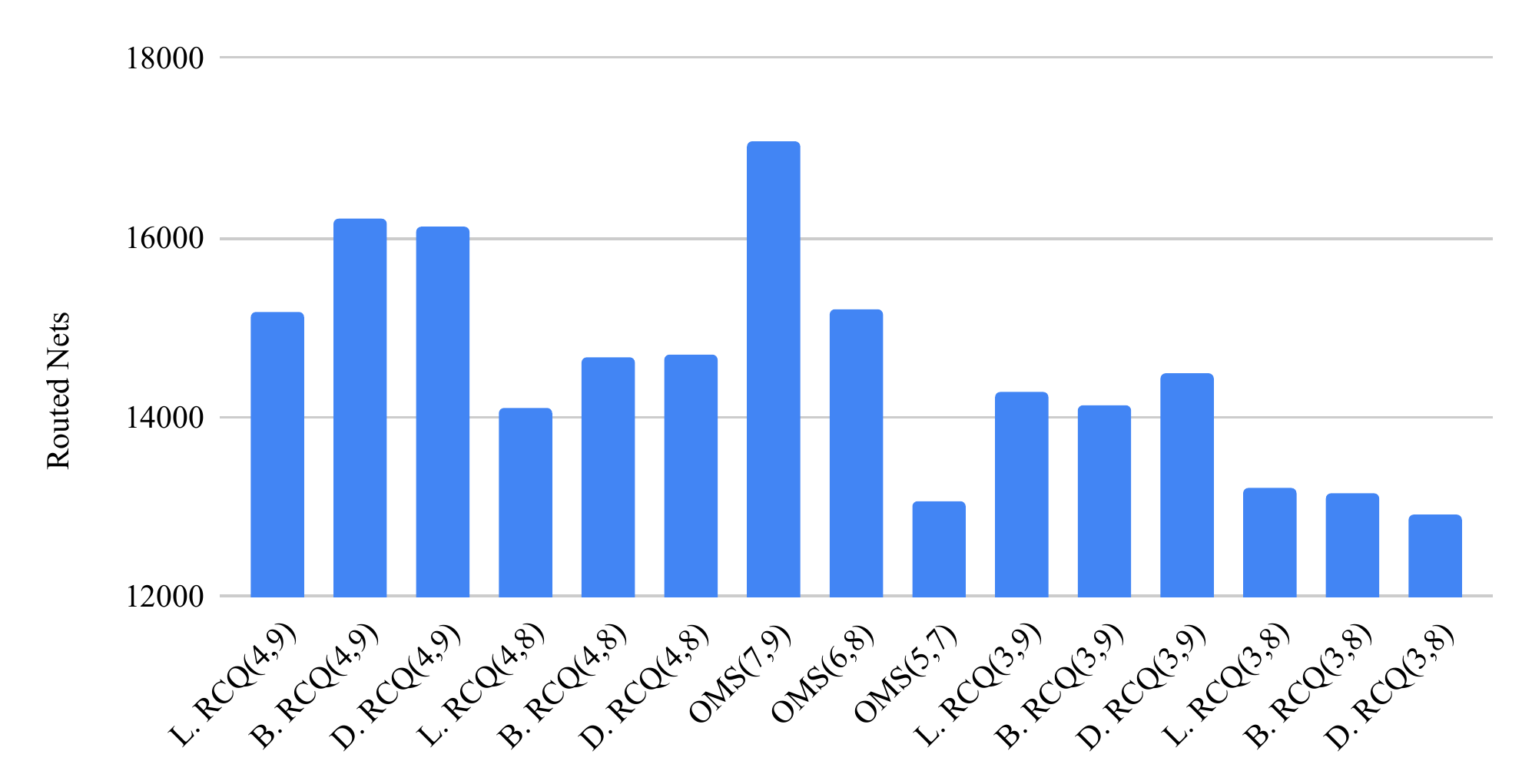}} 
  \caption{ Resource  Utilization of OMS and L-msRCQ Decoders.}
\end{figure}

Fig.\ref{fig: bram_power} shows the BRAM usage and power consumption of each decoder. Power consumption is measured using the Xilinx Vivado Design Suite's power estimation capabilities. Since the static power usage of all the designs is 0.597$W$, we compare the dynamic power consumption. The large amount of L-msRCQ data storage in VN banks for the Lookup-based decoders requires these designs to use by far the most BRAMs. Additionally, designs using $b^v=9$ bits for $V_n$ require more BRAMs as compared to the designs using only $b^v=8$ bits. Fig.\ref{fig: bram_power} reveals that large BRAM usage induces a large power consumption.

\subsection{Routed Nets}

To understand the wiring cost of the L-msRCQ designs, Fig. \ref{fig: rnet} shows the number of routed nets in each architecture. A reduced message size and fewer bits being used in the CN pipeline decreases the number of routes for the 3-bit L-msRCQ designs and also in the RCQ(4,8) designs compared to OMS(6,8). The Broadcast-based L-msRCQ decoders do not experience the expected increase in routing resources used in order to wire $Q(\cdot)$ and $R(\cdot)$ parameters to VN banks. This can be attributed to the decrease of routing elsewhere in the designs such as with message passing and CN operations, effectively offsetting the extra routes for the L-msRCQ data.

\begin{table}[t]
\vspace{0.03in}
\caption{Decoder FER Performance and Resource Usage}
\begin{center}
\begin{tabu}{|c|c|c|c|c|c|c|}
\hline
\textbf{decoder} & {\scriptsize \textbf{Eb/No$^{\mathrm{a}}$}} & \textbf{LUT} & \textbf{Reg.}& {\scriptsize \textbf{BRAM}} & {\scriptsize\textbf{Rt. Net}} & \textbf{Power$^{\mathrm{b}}$} \\
\tabucline[1.25pt]{-}
baseline$^{\mathrm{c}}$ & 1.51 & 9484 & 9518 & 33 & 15201 & 0.757
\\ \tabucline[1.25pt]{-}
{\scriptsize L. RCQ(4,9)} & {\color{mygreen}-0.05} & {\color{mygreen}-11\%} & {\color{mygreen}-3\%} & {\color{red}+170\%} & 0\% & {\color{red}+75\%}  \\
\hline
{\scriptsize B. RCQ(4,9)} & {\color{mygreen}-0.05} & {\color{blue}+10\%} & {\color{mygreen}-4\%} & {\color{blue}+29\%} & {\color{blue}+7\%} & {\color{blue}+15\%}  \\
\hline
{\scriptsize D. RCQ(4,9)} & {\color{mygreen}-0.05} & {\color{blue}+12\%} & {\color{mygreen}-4\%} & {\color{blue}+27\%} & {\color{blue}+6\%} & {\color{blue}+13\%}  \\
\hline
{\scriptsize L. RCQ(4,8)} & {\color{mygreen}-0.03} & {\color{mygreen}-16\%} & {\color{mygreen}-10\%} & {\color{red}+97\%} & {\color{mygreen}-7\%} & {\color{blue}+45\%}  \\
\hline
{\scriptsize B. RCQ(4,8)} & {\color{mygreen}-0.03} & {\color{blue}+1\%} & {\color{mygreen}-12\%} & {\color{blue}+3\%} & {\color{mygreen}-3\%} & 0\%  \\
\hline
{\scriptsize D. RCQ(4,8)} & {\color{mygreen}-0.03} & {\color{blue}+2\%} & {\color{mygreen}-11\%} & {\color{blue}+3\%} & {\color{mygreen}-3\%} & {\color{blue}+2\%}  \\
\hline
{\scriptsize OMS(7,9)} & 0 & {\color{blue}+18\%} & {\color{blue}+12\%} & {\color{blue}+24\%} & {\color{blue}+12\%} & {\color{blue}+18\%}  \\
\hline
{\scriptsize OMS(6,8)} & 0 & 0\% & 0\% & 0\% & 0\% & 0\%  \\
\hline
{\scriptsize OMS(5,7)} & {\color{red}N/A} & {\color{mygreen}-15\%} & {\color{mygreen}-14\%} & 0\% & {\color{mygreen}-14\%} & {\color{mygreen}-8\%}  \\
\hline
{\scriptsize L. RCQ(3,9)} & {\color{blue}+0.06} & {\color{mygreen}-17\%} & {\color{mygreen}-9\%} & {\color{red}+121\%} & {\color{mygreen}-6\%} & {\color{red}+57\%}   \\
\hline
{\scriptsize B. RCQ(3,9)} & {\color{blue}+0.06} & {\color{mygreen}-8\%} & {\color{mygreen}-12\%} & {\color{blue}+27\%} & {\color{mygreen}-7\%} & {\color{blue}+11\%}  \\
\hline
{\scriptsize D. RCQ(3,9)} & {\color{blue}+0.06} & {\color{mygreen}-8\%} & {\color{mygreen}-10\%} & {\color{blue}+27\%} & {\color{mygreen}-5\%} & {\color{blue}+9\%}  \\
\hline
{\scriptsize L. RCQ(3,8)} & {\color{blue}+0.09} & {\color{mygreen}-22\%} & {\color{mygreen}-16\%} & {\color{red}+97\%} & {\color{mygreen}-13\%} & {\color{blue}+41\%}  \\
\hline
{\scriptsize B. RCQ(3,8)} & {\color{blue}+0.09} & {\color{mygreen}-15\%} & {\color{mygreen}-18\%} & {\color{blue}+3\%} & {\color{mygreen}-13\%} & {\color{mygreen}-3\%}  \\
\hline
{\scriptsize D. RCQ(3,8)} & {\color{blue}+0.09} & {\color{mygreen}-17\%} & {\color{mygreen}-17\%} & {\color{blue}+3\%} & {\color{mygreen}-15\%} & {\color{mygreen}-4\%}  \\
\hline
\multicolumn{7}{l}{$^{\mathrm{a}}$Estimated signal to noise ratio $\left ( \frac{E_b}{N_o} \right ) $ to achieve FER of $10^{-7}.$}\\
\multicolumn{7}{l}{$^{\mathrm{b}}$Dynamic power usage measured in watts.}\\
\multicolumn{7}{l}{$^{\mathrm{c}}$FER performance and resource usage of the OMS(6,8) decoder.}
\end{tabu}
\label{tab1}
\end{center}
\end{table}
% Based on the resource and power usage reports, the Broadcast and Dribble methods provide the least-costly implementations of L-msRCQ. Using these techniques for 4-bit L-msRCQ, a 0.05 dB increase in FER performance is able to be achieved with a 13\% reduction in register usage,\textcolor{red}{compared with .. }. Meanwhile for 3-bit L-msRCQ, slight decreases in the LUT, register, wiring, and dynamic power usage are experienced at the cost of a 0.05 dB decrease in FER performance. However, it is to be noted that the OMS(5,7) design achieves better FER performance than the 3-bit L-msRCQ schemes, while using a similar amount of resources as L-msRCQ(3,8). \textcolor{red}{Thus, if minimizing resource utilization is the primary objective for a decoder, then OMS(5,7) would be the better design.} For greater FER performance while using similar amounts of hardware, the error-correction abilities of the 4-bit L-msRCQ designs are unmatched by the OMS decoders.

\subsection{Overall Comparison}

Table \ref{tab1} summarizes the FER performance and resource utilization data of all the implemented decoders compared to the OMS(6,8) decoder, which we treat as a baseline. The OMS(5,7) decoder has an error floor above FER $10^{-5}$, which removes it from consideration as the reference. The OMS(7,9) decoder has an FER performance that is too similar to the OMS(6,8) decoder to justify its increased resource utilization.

The Table \ref{tab1} comparisons confirm that the Lookup-based L-msRCQ decoders require significantly more BRAMs and dynamic power than the other L-msRCQ architectures. Since the Lookup method does not provide any other significant resource advantages, we remove it from consideration. The Broadcast and Dribble architectures provide very similar utilizations for every considered resource. We focus on the Broadcast method in the following analysis due to its simplicity.

The best FER performance is provided by the highest-precision RCQ(4,9) decoders, which perform 0.05 dB better than the baseline OMS(6,8) decoder at FER $10^{-7}$. This performance is costly, using 29\% more BRAMS and 15\% more dynamic power for the Broadcast-based implementation. The Broadcast RCQ(4,8) decoder performs 0.03 dB better than the OMS(6,8) decoder while using 12\% fewer registers than the baseline system and having an otherwise similar resource utilization. The Broadcast RCQ(3,9) decoder performs 0.06 dB worse than the OMS(6,8) decoder at FER $10^{-7}$ and requires 27\% more BRAMS and 11\% more power, but uses 12\% fewer registers. The Broadcast RCQ(3,8) structure performs 0.9 dB worse than the OMS(6,8) decoder, but uses 15\% fewer LUTs, 18\% fewer registers, and 13\% fewer routed nets while requiring very similar amounts of BRAMs and power. 

In summary, L-msRCQ with 8 bits used for VN bank storage is the more preferable architecture. With 4-bit messages and a Broadcast implementation, this architecture provides a slightly better FER than the baseline, while using fewer registers. With 3-bit messages and a Broadcast approach, the FER is slightly worse than OMS(6,8) but allows significant resource reductions in LUTs, registers, and routed nets.

\section{Conclusion}
In this paper, we investigate the FPGA implementations for a L-msRCQ decoding approach. Three L-msRCQ implementations are proposed in this paper: Lookup, Broadcast, and  Dribble. We implemented a layered OMS decoder as a reference and multiple L-msRCQ decoders for a (16384,8192) PBRL code. Simulation results show that with 4-bit messages and a Broadcast implementation, L-msRCQ provides 0.03 dB better FER performance than our baseline OMS decoder and uses 12\% fewer registers. With a Broadcast 3-bit L-msRCQ approach, a 0.09 dB performance loss is experienced, but the design uses 15\% fewer LUTs, 18\% fewer registers, and 13\% fewer routed nets than the OMS decoder.

\bibliographystyle{IEEEtran}

\bibliography{conference_101719.bib}

\begin{thebibliography}{10}
\providecommand{\url}[1]{#1}
\csname url@samestyle\endcsname
\providecommand{\newblock}{\relax}
\providecommand{\bibinfo}[2]{#2}
\providecommand{\BIBentrySTDinterwordspacing}{\spaceskip=0pt\relax}
\providecommand{\BIBentryALTinterwordstretchfactor}{4}
\providecommand{\BIBentryALTinterwordspacing}{\spaceskip=\fontdimen2\font plus
\BIBentryALTinterwordstretchfactor\fontdimen3\font minus
  \fontdimen4\font\relax}
\providecommand{\BIBforeignlanguage}[2]{{%
\expandafter\ifx\csname l@#1\endcsname\relax
\typeout{** WARNING: IEEEtran.bst: No hyphenation pattern has been}%
\typeout{** loaded for the language `#1'. Using the pattern for}%
\typeout{** the default language instead.}%
\else
\language=\csname l@#1\endcsname
\fi
#2}}
\providecommand{\BIBdecl}{\relax}
\BIBdecl

\bibitem{Jinghu_Chen2005-vu}
{Jinghu Chen}, A.~Dholakia, E.~Eleftheriou, M.~P.~C. Fossorier, and {Xiao-Yu
  Hu}, ``Reduced-complexity decoding of {LDPC} codes,'' \emph{IEEE Trans.
  Commun.}, vol.~53, no.~8, pp. 1288--1299, Aug. 2005.

\bibitem{Zhang2007-qu}
Z.~Zhang, L.~Dolecek, M.~Wainwright, V.~Anantharam, and B.~Nikolic,
  ``Quantization effects in {Low-Density} {Parity-Check} decoders,'' in
  \emph{2007 {IEEE} International Conf. on Communications}, Jun. 2007, pp.
  6231--6237.

\bibitem{Stark2019-bm}
M.~{Stark}, G.~{Bauch}, L.~{Wang}, and R.~D. {Wesel}, ``Information bottleneck
  decoding of rate-compatible 5g-ldpc codes,'' \emph{ICC 2020 - 2020 IEEE
  International Conf. on Communications (ICC)}, pp. 1--6, 2020.

\bibitem{Meidlinger2017-aa}
M.~Meidlinger and G.~Matz, ``On irregular {LDPC} codes with quantized message
  passing decoding,'' in \emph{2017 {IEEE} 18th International Workshop on
  Signal Processing Advances in Wireless Communications ({SPAWC})}, Jul. 2017,
  pp. 1--5.

\bibitem{Stark2018-ff}
M.~Stark, J.~Lewandowsky, and G.~Bauch, ``{Information-Optimum} {LDPC} decoders
  with message alignment for irregular codes,'' in \emph{2018 {IEEE} Global
  Communications Conf. ({GLOBECOM})}, Dec. 2018, pp. 1--6.

\bibitem{Lewandowsky2018-pg}
J.~Lewandowsky and G.~Bauch, ``{Information-Optimum} {LDPC} decoders based on
  the information bottleneck method,'' \emph{IEEE Access}, vol.~6, pp.
  4054--4071, 2018.

\bibitem{Romero2016-jn}
F.~J.~C. Romero and B.~M. Kurkoski, ``{LDPC} decoding mappings that maximize
  mutual information,'' \emph{IEEE J. Sel. Areas Commun.}, vol.~34, no.~9, pp.
  2391--2401, Sep. 2016.

\bibitem{-_Lee2005-ni}
J.~K.~S. {Lee} and J.~{Thorpe}, ``Memory-efficient decoding of ldpc codes,'' in
  \emph{Proceedings. International Symposium on Information Theory, 2005. ISIT
  2005.}, 2005, pp. 459--463.

\bibitem{He2019-hr}
X.~He, K.~Cai, and Z.~Mei, ``On mutual {Information-Maximizing} quantized
  belief propagation decoding of {LDPC} codes,'' in \emph{2019 {IEEE} Global
  Communications Conf. ({GLOBECOM})}, Dec. 2019, pp. 1--6.

\bibitem{Wang2020-kh}
L.~Wang, R.~D. Wesel, M.~Stark, and G.~Bauch, ``A
  {Reconstruction-Computation-Quantization} ({RCQ}) approach to node operations
  in {LDPC} decoding,'' in \emph{{GLOBECOM} 2020 - 2020 {IEEE} Global
  Communications Conf.}, Dec. 2020, pp. 1--6.

\bibitem{Luby2001-ut}
M.~G. Luby, M.~Mitzenmacher, M.~A. Shokrollahi, and D.~A. Spielman, ``Improved
  low-density parity-check codes using irregular graphs,'' \emph{IEEE Trans.
  Inf. Theory}, vol.~47, no.~2, pp. 585--598, Feb. 2001.

\bibitem{-_Chang2008-vk}
Y.~.~Chang, A.~I.~V. Casado, M.~.~F.~Chang, and R.~D. Wesel,
  ``{Lower-Complexity} layered {Belief-Propagation} decoding of {LDPC} codes,''
  in \emph{2008 {IEEE} International Conf. on Communications}, May 2008, pp.
  1155--1160.

\bibitem{Chen2015-ul}
T.~Chen, K.~Vakilinia, D.~Divsalar, and R.~D. Wesel, ``{Protograph-Based}
  {Raptor-Like} {LDPC} codes,'' \emph{IEEE Trans. Commun.}, vol.~63, no.~5, pp.
  1522--1532, May 2015.

\bibitem{Fossorier1999-bp}
M.~P.~C. Fossorier, M.~Mihaljevic, and H.~Imai, ``Reduced complexity iterative
  decoding of low-density parity check codes based on belief propagation,''
  \emph{IEEE Trans. Commun.}, vol.~47, no.~5, pp. 673--680, May 1999.

\bibitem{Chen2002-xi}
J.~Chen and M.~P.~C. Fossorier, ``Density evolution for two improved {BP-Based}
  decoding algorithms of {LDPC} codes,'' \emph{IEEE Commun. Lett.}, vol.~6,
  no.~5, pp. 208--210, May 2002.

\bibitem{Hocevar2004-xi}
D.~E. Hocevar, ``A reduced complexity decoder architecture via layered decoding
  of {LDPC} codes,'' in \emph{{IEEE} Workshop onSignal Processing Systems,
  2004. {SIPS} 2004.}, Oct. 2004, pp. 107--112.

\end{thebibliography}

\end{document}